\documentclass[aps,nofootinbib,superscriptaddress]{revtex4}%
\usepackage{hyperref}
\usepackage{amsmath}
\usepackage{amsfonts}
\usepackage{amssymb}
\usepackage{graphicx}
\usepackage{color}%
\providecommand{\U}[1]{\protect\rule{.1in}{.1in}}

\begin{document}
\title{Stringy Scaling\smallskip\ }
\author{Sheng-Hong Lai}
\email{xgcj944137@gmail.com}
\affiliation{Department of Electrophysics, National Yang Ming Chiao Tung University,
Hsinchu, Taiwan, R.O.C.}
\author{Jen-Chi Lee}
\email{jcclee@cc.nctu.edu.tw}
\affiliation{Department of Electrophysics, National Yang Ming Chiao Tung University,
Hsinchu, Taiwan, R.O.C.}
\author{Yi Yang}
\email{yiyang@mail.nctu.edu.tw}
\affiliation{Department of Electrophysics, National Yang Ming Chiao Tung University,
Hsinchu, Taiwan, R.O.C.}

\begin{abstract}
We discover a general \textit{stringy scaling} behavior for\medskip

1. \textbf{All n-point} \textbf{hard} string scattering amplitudes ($HSSA$)
and\medskip

2. A class of $n$-point \textbf{Regge} string scattering amplitudes
($RSSA$)\smallskip\medskip

to all string loop orders.\medskip

The number of independent kinematics variables is found to be reduced by
dim$\mathcal{M}$.

\end{abstract}
\date{\today}
\maketitle

\section{The first example of stringy scaling}

\textbf{FACT} : All $4$-point HSSA ($E\rightarrow\infty$, fixed $\phi$) at
each fixed mass level of $V_{2}$ (with three other vertices $V_{1},V_{3}$ and
$V_{4}$ fixed) share the same functional form.

That is, all HSSA at each fixed mass level are proportional to each other with
\textit{constant} \textbf{ratios} (independent of the scattering angle $\phi$).

\begin{figure}[ptb]
\label{scattering} \setlength{\unitlength}{3pt}
\par
\begin{center}
\begin{picture}(100,70)(-50,-30)
{\large
\put(45,0){\vector(-1,0){42}} \put(-45,0){\vector(1,0){42}}
\put(2,2){\vector(1,1){30}} \put(-2,-2){\vector(-1,-1){30}}
\put(25,2){$k_1$} \put(-27,2){$k_2$} \put(11,20){$-k_3$}
\put(-24,-15){$-k_4$}
\put(40,0){\vector(0,-1){10}} \put(-40,0){\vector(0,1){10}}
\put(26,26){\vector(-1,1){7}} \put(-26,-26){\vector(1,-1){7}}
\put(36,-16){$e^{T}(1)$} \put(-44,15){$e^{T}(2)$}
\put(15,36){$e^{T}(3)$} \put(-18,-35){$e^{T}(4)$}
\qbezier(10,0)(10,4)(6,6) \put(12,4){$\phi$}
}
\end{picture}
\end{center}
\caption{Kinematic variables in the center of mass frame}%
\end{figure}
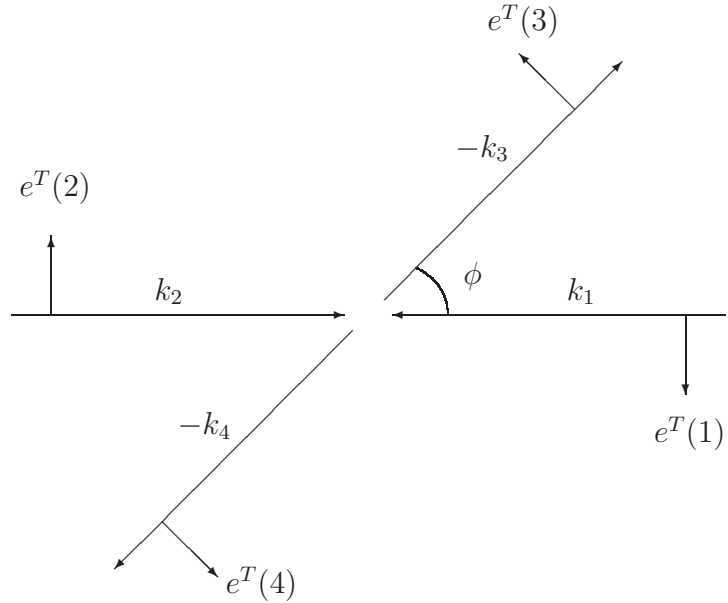

\bigskip\textbf{PROOF} : The starting point is to apply the $4$-point $l$-loop
stringy \textbf{on-shell Ward identities}%
\begin{equation}
\left\langle V_{1}\chi V_{3}V_{4}\right\rangle _{l-loop}=0
\end{equation}
\textbf{in the hard scattering limit}.

$V_{j}$ above can be \textbf{any} string vertex and the second vertex $\chi$
is the vertex of a \textbf{zero-norm state} (ZNS).

In the hard scattering limit, components of polarization orthogonal to the
scattering plane are subleading order in energy.

Defining $e^{P}=\frac{1}{M_{2}}(E_{2},\mathrm{k}_{2},\vec{0})=\frac{k_{2}%
}{M_{2}}$ the \textbf{momentum polarization}, $e^{L}=\frac{1}{M_{2}%
}(\mathrm{k}_{2},E_{2},\vec{0})$ the \textbf{longitudinal polarization} and
the \textbf{transverse polarization} $e^{T}=(0,0,1)$ on the scattering plane.

In the hard scattering limit, it can be shown that at \textbf{each fixed mass
level} $M^{2}=2(N-1)$ only states of the following form (in the hard
scattering limit $e^{P}\simeq$ $e^{L}$)
\begin{equation}
\left\vert N,2m,q\right\rangle =\left(  \alpha_{-1}^{T}\right)  ^{N-2m-2q}%
\left(  \alpha_{-1}^{L}\right)  ^{2m}\left(  \alpha_{-2}^{L}\right)
^{q}\left\vert 0;k\right\rangle \label{11}%
\end{equation}
are leading order in energy.

There are two types of physical ZNS in the \textbf{old covariant} first
quantized open bosonic string spectrum: \cite{GSW}
\begin{equation}
\text{Type I}:L_{-1}\left\vert y\right\rangle ,\text{ where }L_{1}\left\vert
y\right\rangle =L_{2}\left\vert y\right\rangle =0,\text{ }L_{0}\left\vert
y\right\rangle =0;
\end{equation}%
\begin{equation}
\text{Type II}:(L_{-2}+\frac{3}{2}L_{-1}^{2})\left\vert \widetilde{y}%
\right\rangle ,\text{ where }L_{1}\left\vert \widetilde{y}\right\rangle
=L_{2}\left\vert \widetilde{y}\right\rangle =0,\text{ }(L_{0}+1)\left\vert
\widetilde{y}\right\rangle =0.\text{(}D=26\text{ \textbf{only})}%
\end{equation}
\qquad

\textbf{(A)} Consider $\chi$ to be the type I hard ZNS (HZNS) calculated from
\textbf{Type I} ZNS
\begin{align}
L_{-1}|N-1,2m-1,q\rangle &  =(M\alpha_{-1}^{L}+\alpha_{-2}^{L}\alpha_{1}%
^{L}+\underset{irrelevant}{\underbrace{\alpha_{-2}^{T}\alpha_{1}^{T}%
+\alpha_{-3}\cdot\alpha_{2}+\cdots}})|N-1,2m-1,q\rangle\\
&  \simeq M|N,2m,q\rangle+(2m-1)|N,2m-2,q+1\rangle
\end{align}
where many terms are omitted because they are not of the form of Eq.(\ref{11}).

This implies the following relation among $4$-point amplitudes
\begin{equation}
\mathcal{T}^{(N,2m,q)}=-\frac{2m-1}{M}\mathcal{T}^{(N,2m-2,q+1)}.
\end{equation}
Using this relation repeatedly, we get
\begin{equation}
\mathcal{T}^{(N,2m,q)}=\frac{(2m-1)!!}{(-M)^{m}}\mathcal{T}^{(N,0,m+q)}.
\label{L1}%
\end{equation}

\textbf{(B)} Next, consider another class of HZNS calculated from \textbf{type
II} ZNS
\begin{align}
L_{-2}|N-2,0,q\rangle &  =(\frac{1}{2}\alpha_{-1}^{T}\alpha_{-1}^{T}%
+M\alpha_{-2}^{L}+\underset{irrelevant}{\underbrace{\alpha_{-3}\cdot\alpha
_{1}+\cdots}})|N-2,0,q\rangle\\
&  \simeq\frac{1}{2}|N,0,q\rangle+M|N,0,q+1\rangle.
\end{align}
Again, irrelevant terms are omitted here. From this we deduce that
\begin{equation}
\mathcal{T}^{(N,0,q+1)}=-\frac{1}{2M}\mathcal{T}^{(N,0,q)},
\end{equation}
which leads to
\begin{equation}
\mathcal{T}^{(N,0,q)}=\frac{1}{(-2M)^{q}}\mathcal{T}^{(N,0,0)}. \label{L2}%
\end{equation}

\textbf{In conclusion}, the decoupling of ZNS Eq.(\ref{L1}) and Eq.(\ref{L2})
leads to\textbf{ ratios} among $4$-point $HSSA$.

(\textbf{Conjectured} by Gross \cite{Gross}, \textbf{proved} by
\cite{ChanLee,ChanLee2,CHLTY2,CHLTY1})%
\begin{equation}
\frac{\mathcal{T}^{\left(  N,m,q\right)  }}{\mathcal{T}^{\left(  N,0,0\right)
}}=\frac{\left(  2m\right)  !}{m!}\left(  \frac{-1}{2M}\right)  ^{2m+q}%
.\text{(\textbf{independent of }}\phi\text{ !!).} \label{22}%
\end{equation}
In Eq.(\ref{22}) $\mathcal{T}^{\left(  N,m,q\right)  }$ is the $4$-point
$HSSA$ of any string vertex $V_{j}$ with $j=1,3,4$ and $V_{2}$ is the high
energy state in Eq.(\ref{11}); while $\mathcal{T}^{\left(  N,0,0\right)  }$ is
the $4$-point $HSSA$ of any string vertex $V_{j}$ with $j=1,3,4$, and $V_{2}$
is the leading Regge trajectory string state at mass level $N$. Note that we
have omitted the tensor indice of $V_{j}$ with $j=1,3,4$ and keep only those
of $V_{2}$ in $\mathcal{T}^{\left(  N,2m,q\right)  }$. \textbf{END}

\subsection{Examples}

Since the ratios are independent of the choices of $V_{1}$, $V_{3}$ and
$V_{4}$, we choose them to be tachyons and $V_{2}$ to be Eq.(\ref{11}).

Since the ratios are independent of the loop order, we choose to calculate
$l=0$ loop.

An explicit amplitude calculation gives \cite{review,over}

\textbf{For} $M^{2}=4$%
\begin{equation}
\mathcal{T}_{TTT}:\mathcal{T}_{(LLT)}:\mathcal{T}_{(LT)}:\mathcal{T}%
_{[LT]}=8:1:-1:-1.\nonumber
\end{equation}

\textbf{For} $M^{2}=6$%
\begin{align}
&  \mathcal{T}_{(TTTT)}:\mathcal{T}_{(TTLL)}:\mathcal{T}_{(LLLL)}%
:\mathcal{T}_{TT,L}:\mathcal{T}_{(TTL)}:\mathcal{T}_{(LLL)}:\mathcal{T}%
_{(LL)}\nonumber\\
&  =16:\frac{4}{3}:\frac{1}{3}:-\frac{2\sqrt{6}}{3}:-\frac{4\sqrt{6}}%
{9}:-\frac{\sqrt{6}}{9}:\frac{2}{3}. \label{61}%
\end{align}
\textbf{For} $M^{2}=8$.%
\begin{align}
&  \mathcal{T}_{(TTTTT)}:\mathcal{T}_{(TTTL)}:\mathcal{T}_{(TTTLL)}%
:\mathcal{T}_{(TLLL)}:\mathcal{T}_{(TLLLL)}:\mathcal{T}_{(TLL)}:\mathcal{T}%
_{T,LL}:\mathcal{T}_{TLL,L}:\mathcal{T}_{TTT,L}\nonumber\\
&  =32:\sqrt{2}:2:\frac{3\sqrt{2}}{16}:\frac{3}{8}:\frac{1}{3}:\frac{2}%
{3}:\frac{\sqrt{2}}{16}:3\sqrt{2}. \label{mass8}%
\end{align}
These are all \textbf{remarkably} consistent with Eq.(\ref{22}) of ZNS calculation.

\textbf{For} \textbf{subleading order amplitudes}, they are in general
\textbf{NOT} proportional to each other.

\textbf{For} $M^{2}=4$, for example, one has $6$ \textbf{subleading order
amplitudes} and $4$ linear relations (on-shell Ward identities) in the ZNS calculation.

An explicit subleading order amplitude calculation gives \cite{review,over}%

\begin{equation}
\mathcal{T}_{LLL}^{2}=-4E^{8}\sin\phi\cos\phi\mathcal{T}(3),
\end{equation}

\begin{equation}
\mathcal{T}_{LTT}^{2}=-8E^{8}\sin^{2}\phi\cos\phi\mathcal{T}(3),
\end{equation}
which show that the proportional coefficients do depend on the scattering
angle $\phi$.

\textbf{IN FIELD THEORY}, as \textbf{an example}, the leading order process of
the elastic scattering of a spin-one-half particle by a spin-zero particle
such as $e^{-}\pi^{+}\longrightarrow e^{-}\pi^{+}$ .

The non-vanishing amplitudes were shown to be%
\begin{align}
\mathcal{T}\text{ }(e_{R}^{-}\pi^{+}  &  \longrightarrow e_{R}^{-}\pi
^{+})=\mathcal{T}\text{ }(e_{L}^{-}\pi^{+}\longrightarrow e_{L}^{-}\pi
^{+})\sim\text{ }\cos\frac{\phi}{2},\\
\mathcal{T}\text{ }(e_{R}^{-}\pi^{+}  &  \longrightarrow e_{L}^{-}\pi
^{+})=\mathcal{T}\text{ }(e_{L}^{-}\pi^{+}\longrightarrow e_{R}^{-}\pi
^{+})\sim\text{ }\sin\frac{\phi}{2}%
\end{align}
which are \textbf{NOT} proportional to each other.

In QED, as \textbf{another example}, for the leading order process of
$e^{-}e^{+}\longrightarrow\mu^{-}\mu^{+}$, there are $4$ non-vanishing among
$16$ hard polarized amplitudes%
\begin{align}
\mathcal{T}\text{ }(e_{R}^{-}e_{L}^{+}  &  \longrightarrow\mu_{R}^{-}\mu
_{L}^{+})=\mathcal{T}\text{ }(e_{L}^{-}e_{R}^{+}\longrightarrow\mu_{L}^{-}%
\mu_{R}^{+})\sim\text{ }(1+\cos\theta)=2\text{ }\cos^{2}\frac{\phi}{2},\\
\mathcal{T}\text{ }(e_{R}^{-}e_{L}^{+}  &  \longrightarrow\mu_{L}^{-}\mu
_{R}^{+})=\mathcal{T}\text{ }(e_{L}^{-}e_{R}^{+}\longrightarrow\mu_{R}^{-}%
\mu_{L}^{+})\sim\text{ }(1-\cos\theta)=2\text{ }\sin^{2}\frac{\phi}{2},
\end{align}
and they are \textbf{NOT} all proportional to each other.

\section{Stringy scaling of higher point ($n\geq5$) $HSSA$}

The stringy \textbf{on-shell Ward identities} can be written as%
\begin{equation}
\left\langle V_{1}\chi V_{3}\cdots V_{n}\right\rangle _{l-loop}=0 \label{w}%
\end{equation}
where $\chi$ again is the vertex of a ZNS.

\subsection{On the scattering plane}

In the hard scattering limit \textbf{on the scattering plane}, the space part
of momenta $k_{j}$ ( $j=3,4,\cdots,n$) form a closed $1$-chain with $(n-2)$
sides due to momentum conservation.

All $4$-point calculation above persist and one ends up with Eq.(\ref{22}) again.

\textbf{HOWEVER}, while for $n=4$, the \textit{ratios} are independent of $1$
scattering angle $\phi$,

\textbf{for} $n=5$, the ratios are independent of $3$ kinematics variables
($2$ angles and $1$ fixed ratio of two infinite energies) or, for simplicity,
$3$ scattering "angles".

\textbf{For} $n=6$, there are $5$ scattering "angles"$\cdots$.

\subsection{Out of the scattering plane}

The general high energy states at each fixed mass level $M^{2}=2(N-1)$ can be
written as%
\begin{equation}
\left\vert \left\{  p_{i}\right\}  ,2m,2q\right\rangle =\left(  \alpha
_{-1}^{T_{1}}\right)  ^{N+p_{1}}\left(  \alpha_{-1}^{T_{2}}\right)  ^{p_{2}%
}\cdots\left(  \alpha_{-1}^{T_{r}}\right)  ^{p_{r}}\left(  \alpha_{-1}%
^{L}\right)  ^{2m}\left(  \alpha_{-2}^{L}\right)  ^{q}\left\vert
0;k\right\rangle
\end{equation}
where $\sum_{i=1}^{r}p_{i}=-2(m+q)$ with\ $r\leq24$.

One generalizes the transverse polarization $e^{T}=(0,0,1)$ to $e^{\hat{T}%
}=(0,0,\vec{\omega})$ where%
\begin{equation}
\omega_{i}=\cos\theta_{i}\prod\limits_{\sigma=1}^{i-1}\sin\theta_{\sigma
}\text{, with }i=1,\cdots,r,\text{ }\theta_{r}=0 \label{ww}%
\end{equation}
are the \textbf{solid angles} in the \textbf{transverse space} spanned by $24$
transverse directions $e^{T_{i}}$.

Note that $\alpha_{-1}^{\hat{T}}=\alpha_{-1}\cdot e^{\hat{T}}$ etc. With
$\left(  \alpha_{-1}^{T_{i}}\right)  =\left(  \alpha_{-1}^{\hat{T}}\right)
\omega_{i}$, we easily obtain%
\begin{align}
&  \left(  \alpha_{-1}^{T_{1}}\right)  ^{N+p_{1}}\left(  \alpha_{-1}^{T_{2}%
}\right)  ^{p_{2}}\cdots\left(  \alpha_{-1}^{T_{r}}\right)  ^{p_{r}}\left(
\alpha_{-1}^{L}\right)  ^{2m}\left(  \alpha_{-2}^{L}\right)  ^{q}\left\vert
0;k\right\rangle \nonumber\\
&  =\left(  \omega_{1}^{N}\prod_{i=1}^{r}\omega_{i}^{p_{i}}\right)  \left(
\alpha_{-1}^{\hat{T}}\right)  ^{N-2m-2q}\left(  \alpha_{-1}^{L}\right)
^{2m}\left(  \alpha_{-2}^{L}\right)  ^{q}\left\vert 0;k\right\rangle ,
\end{align}
which leads to the ratios of $n$-point $HSSA$%
\begin{equation}
\frac{\mathcal{T}^{\left(  \left\{  p_{i}\right\}  ,2m,2q\right)  }%
}{\mathcal{T}^{\left(  \left\{  0_{i}\right\}  ,0,0\right)  }}=\frac{\left(
2m\right)  !}{m!}\left(  \frac{-1}{2M}\right)  ^{2m+q}\prod_{i=1}^{r}%
\omega_{i}^{p_{i}}\text{ (\textbf{\# of kinematic variables dependence
reduced} !!)} \label{100}%
\end{equation}
where $\mathcal{T}^{\left(  \left\{  0_{i}\right\}  ,0,0\right)  }$ is the
$HSSA$ of leading \textbf{Regge trajectory state} at mass level $M^{2}=2(N-1)$.

\textbf{These ratios are valid to all string loop orders.}

\subsection{Stringy Scaling}

\textbf{For} the simple case with $n=4$ and $r=1$, one has two variables, $s$
and $t$ (or $E$, $\phi$).

The ratios of all HSSA are independent of the scattering angle $\phi$ and
dim$\mathcal{M}=1$.

\textbf{For} the general $n$-point $HSSA$ with $r\leq24$, $d=r+2$, we have
$k_{j}$ vector with $j=1,\cdots,n$ and $k_{j}$ $\in R^{d-1,1}$.

The number of kinematics variables is $n\left(  d-1\right)  -\frac{d\left(
d+1\right)  }{2}$.

Indeed, as $p=E\rightarrow\infty$, we define the $26$-dimensional momenta in
the CM frame to be%
\begin{align}
k_{1}  &  =\left(  E,-E,0^{r}\right)  ,\nonumber\\
k_{2}  &  =\left(  E,+E,0^{r}\right)  ,\nonumber\\
k_{j}  &  =\left(  -q_{j},-q_{j}\Omega_{1}^{j},\cdots-q_{j}\Omega_{25}%
^{j}\right)  \label{k22}%
\end{align}
where $j=3,4,\cdots,n$, and%
\begin{equation}
\Omega_{i}^{j}=\cos\phi_{i}^{j}\prod\limits_{\sigma=1}^{i-1}\sin\phi_{\sigma
}^{j}\text{ with }\phi_{j-1}^{j}=0,\text{ }\phi_{i>r}^{j}=0\text{ and }%
r\leq\min\left\{  n-3,24\right\}  \label{k33}%
\end{equation}
are the solid angles in the $\left(  j-2\right)  $-dimensional spherical space
with $\sum_{i=1}^{j-2}\left(  \Omega_{i}^{j}\right)  ^{2}=1$. In
Eq.(\ref{k22}), $0^{r}$ denotes the $r$-dimensional null vector. The condition
$\phi_{j-1}^{j}=0$ in Eq.(\ref{k33}) was chosen to fix the frame by using the
rotational symmetry. The independent kinematics variables can be chosen to be
some $\varphi_{i}^{j}$ and some fixed ratios of infinite $q_{j}$.

For the kinematics parameter space $\mathcal{M}$ defined by
\begin{equation}
\omega_{j}\left(  \text{kinematics parameters with }E\rightarrow\infty\right)
=\text{fixed constant \ }(j=2,\cdots,r)\text{,}%
\end{equation}
we can count the dimension of $\mathcal{M}$ to be
\begin{equation}
\text{dim}\mathcal{M}\text{ }\mathcal{=}\text{ }n\left(  d-1\right)
-\frac{d\left(  d+1\right)  }{2}-1-\left(  r-1\right)  =\frac{\left(
r+1\right)  \left(  2n-r-6\right)  }{2}%
\end{equation}
where $r=d-2$ is the number of transverse directions $e^{T_{i}}$.

\textbf{For} $n=5$ and $r=2$, as an example, $d=r+2=4$ and one has $n\left(
d-1\right)  -\frac{d\left(  d+1\right)  }{2}=5$ parameters ($r_{1}$ is the
ratio of two infinite energies)%
\begin{equation}
E,\phi_{2}^{3},\phi_{2}^{4},\phi_{3}^{4},r_{1}.
\end{equation}
In the hard scattering limit $E\rightarrow\infty$, for $\theta_{1}=fixed$ we
get dim$\mathcal{M=}$ $3$.

\textbf{For} $n=6$ and $r=3$, as another example, the ratios of $6$-point
$HSSA$ depends only on $2$ variables $\theta_{1}$ and $\theta_{2}$ instead of
$8$ "angles" and dim$\mathcal{M=}$ $6$.

\textbf{In sum}, the ratios among $n$-point $HSSA$ with $r\leq24$ are
constants and independent of the scattering "angles" in the kinematic regime
$\mathcal{M}$.

\subsection{Example}

For $n=6$ and $r=3$, as an example, $\mathcal{M}$ is defined by
\begin{equation}
\theta_{j}\left(  8\text{ kinematics parameters}\right)  =\text{fixed
constant, \ }j=1,2,
\end{equation}
and we have $\dim\mathcal{M=}$ $6$.

For this case, the \textbf{ratios}%
\begin{equation}
\frac{\mathcal{T}^{\left(  \left\{  p_{1},p_{2},p_{3}\right\}  ,m,q\right)  }%
}{\mathcal{T}^{\left(  \left\{  0,0,0\right\}  ,0,0\right)  }}=\frac{\left(
2m\right)  !}{m!}\left(  \frac{-1}{2M}\right)  ^{2m+q}\left(  \cos\theta
_{1}\right)  ^{p_{1}}\left(  \sin\theta_{1}\cos\theta_{2}\right)  ^{p_{2}%
}\left(  \sin\theta_{1}\sin\theta_{2}\right)  ^{p_{3}} \label{16}%
\end{equation}
are independent of kinematics parameters in the space $\mathcal{M}$.

\textbf{For example}, for say $\theta_{1}=\frac{\pi}{4}$ and $\theta_{2}%
=\frac{\pi}{6}$, we get the \textbf{ratios among }$6$\textbf{-point }$HSSA$%
\begin{equation}
\frac{\mathcal{T}^{\left(  \left\{  p_{1},p_{2},p_{3}\right\}  ,m,q\right)  }%
}{\mathcal{T}^{\left(  \left\{  0,0,0\right\}  ,0,0\right)  }}=\left(
-\frac{1}{M}\right)  ^{2m+q}(2m-1)!!\left(  \frac{1}{2}\right)  ^{p_{2}+p_{3}%
}\left(  \sqrt{3}\right)  ^{p_{3}}\text{.}%
\end{equation}

\subsection{Summary}

In the hard scattering limit, the number of scattering "angles" dependence on
ratios of $n$-point $HSSA$ with $r\leq24$ \textbf{reduces} by dim$M$.

For a given ($n,r$)%
\[%
\begin{tabular}
[c]{|c|c|c|c|c|}\hline
$\text{dim}\mathcal{M}$ & $r=1$ & $r=2$ & $r=3$ & $r=4$\\\hline
$n=4$ & $1$ &  &  & \\\hline
$n=5$ & $3$ & $3$ &  & \\\hline
$n=6$ & $5$ & $6$ & $6$ & \\\hline
$n=7$ & $7$ & $9$ & $10$ & $10$\\\hline
\end{tabular}
\
\]

\bigskip

\section{Saddle point calculation}

We begin with the $4$-point case \cite{CHLTY2,CHLTY1}

\subsection{The four point calculation}

Since the ratios are independent of the choices of $V_{1}$, $V_{3}$ and
$V_{4}$, we choose them to be tachyons and $V_{2}$ to be Eq.(\ref{11}).

Since the ratios are independent of the loop order, we choose to calculate
$l=0$ loop.

The $t-u$ channel contribution to the stringy amplitude at tree level is
(after $SL(2,R)$ fixing)%
\begin{align}
\mathcal{T}^{(N,2m,q)}  &  =\int_{1}^{\infty}dxx^{(1,2)}(1-x)^{(2,3)}\left[
\frac{e^{T}\cdot k_{1}}{x}-\frac{e^{T}\cdot k_{3}}{1-x}\right]  ^{N-2m-2q}%
\nonumber\\
&  \cdot\left[  \frac{e^{P}\cdot k_{1}}{x}-\frac{e^{P}\cdot k_{3}}%
{1-x}\right]  ^{2m}\left[  -\frac{e^{P}\cdot k_{1}}{x^{2}}-\frac{e^{P}\cdot
k_{3}}{(1-x)^{2}}\right]  ^{q}%
\end{align}
where $(1,2)=k_{1}\cdot k_{2}$ etc.

In order to apply the \textbf{saddle-point method}, we rewrite the amplitude
above into the following form%
\begin{equation}
\mathcal{T}^{(N,2m,q)}(K)=\int_{1}^{\infty}dx\mbox{ }u(x)e^{-Kf(x)},
\end{equation}
where
\begin{align}
K  &  \equiv-(1,2)\rightarrow\frac{s}{2}\rightarrow2E^{2},\\
\tau &  \equiv-\frac{(2,3)}{(1,2)}\rightarrow-\frac{t}{s}\rightarrow\sin
^{2}\frac{\phi}{2},\\
f(x)  &  \equiv\ln x-\tau\ln(1-x),\\
u(x)  &  \equiv\left[  \frac{(1,2)}{M}\right]  ^{2m+q}(1-x)^{-N+2m+2q}%
\underset{\ast}{\underbrace{(f^{\prime})^{2m}}}(f^{\prime\prime})^{q}%
(-e^{T}\cdot k_{3})^{N-2m-2q}.
\end{align}
The saddle-point for the integration of moduli, $x=x_{0}$, is defined by
\begin{equation}
f^{\prime}(x_{0})=0,
\end{equation}
and we have%
\begin{equation}
x_{0}=\frac{1}{1-\tau}=\sec^{2}\frac{\phi}{2},\hspace{1cm}1-x_{0}=-\frac{\tau
}{1-\tau},\hspace{1cm}f^{\prime\prime}(x_{0})=(1-\tau)^{3}\tau^{-1}.
\label{saddle}%
\end{equation}
It is easy to see that%
\begin{equation}
u(x_{0})=u^{\prime}(x_{0})=....=u^{(2m-1)}(x_{0})=0,
\end{equation}
and
\begin{equation}
u^{(2m)}(x_{0})=\left[  \frac{(1,2)}{M}\right]  ^{2m+q}(1-x_{0})^{-N+2m+2q}%
(2m)!(f_{0}^{\prime\prime})^{2m+q}(-e^{T}\cdot k_{3})^{N-2m-2q}.
\end{equation}

With these inputs, one can easily evaluate the Gaussian integral associated
with the four-point amplitudes
\begin{align}
&  \int_{1}^{\infty}dx\mbox{ }u(x)e^{-Kf(x)}\nonumber\\
&  =\sqrt{\frac{2\pi}{Kf_{0}^{\prime\prime}}}e^{-Kf_{0}}\left[  \frac
{u_{0}^{(2m)}}{2^{m}\ m!\ (f_{0}^{\prime\prime})^{m}\ K^{m}}+O(\frac
{1}{K^{m+1}})\right] \nonumber\\
&  =\sqrt{\frac{2\pi}{Kf_{0}^{\prime\prime}}}e^{-Kf_{0}}\left[  (-1)^{N-q}%
\frac{2^{N-2m-q}(2m)!}{m!\ {M}^{2m+q}}\ \tau^{-\frac{N}{2}}(1-\tau)^{\frac
{3N}{2}}E^{N}+O(E^{N-2})\right]  .
\end{align}
This result shows explicitly that with one tensor and three tachyons, the
energy and angle dependence for the four-point HSS amplitudes only depend on
the level $N$
\begin{align}
\lim_{E\rightarrow\infty}\frac{\mathcal{T}^{(N,2m,q)}}{\mathcal{T}^{(N,0,0)}}
&  =\frac{(-1)^{q}(2m)!}{m!(2M)^{2m+q}}\nonumber\\
&  =(-\frac{2m-1}{M})....(-\frac{3}{M})(-\frac{1}{M})(-\frac{1}{2M})^{m+q},
\end{align}
which is\textbf{ remarkably }consistent with calculation of decoupling of high
energy ZNS obtained in Eq.(\ref{22}).

\bigskip

\subsection{The $n$-point $HSSA$ with $r=1$}

To illustrate the $n$-point $HSSA$ calculation, we begin with $n$-point $HSSA$
with $r=1$.

We want to calculate $n$-point $HSSA$ with $(n-1)$ tachyons and $1$ high
energy state%
\begin{equation}
\left(  \alpha_{-1}^{T}\right)  ^{N-2m-2q}\left(  \alpha_{-1}^{L}\right)
^{2m}\left(  \alpha_{-2}^{L}\right)  ^{q}\left\vert 0;k\right\rangle ,\text{
\ \ }M^{2}=2\left(  N-1\right)  .
\end{equation}
With the \textbf{change of variables} $z_{i}=\frac{x_{i}}{x_{i+1}}$ or
$x_{i}=z_{i}\cdots z_{n-2}$, the $HSSA$ can be written as%
\begin{align}
\mathcal{T}^{\left(  \left\{  p_{i}\right\}  ,m,q\right)  }  &  =\int_{0}%
^{1}dx_{n-2}\cdots\text{ }\int_{0}^{x_{4}}dx_{3}\int_{0}^{x_{3}}dx_{2}%
ue^{-Kf}\nonumber\\
&  =\int_{0}^{1}dz_{n-2}\cdots\text{ }\int_{0}^{1}dz_{3}\int_{0}^{1}dz_{2}%
\begin{vmatrix}
z_{3}\cdots z_{n-2} & z_{2}z_{4}\cdots z_{n-2} & \cdots & z_{2}\cdots
z_{n-3}\\
0 & z_{4}\cdots z_{n-2} & \cdots & \\
&  & \ddots & \\
0 & 0 & \cdots & 1
\end{vmatrix}
ue^{-Kf}\nonumber\\
&  =\left(  \prod_{i=3}^{n-2}\int_{0}^{1}dz_{i}\text{ }z_{i}^{i-2-N}\right)
\int_{0}^{1}dz_{2}ue^{-Kf}%
\end{align}
where%
\begin{align}
f\left(  x_{i}\right)   &  =-\underset{i<j}{\sum}\frac{k_{i}\cdot k_{j}}{K}%
\ln\left(  x_{j}-x_{i}\right)  =-\underset{i<j}{\sum}\frac{k_{i}\cdot k_{j}%
}{K}\ln\left(  z_{j}\cdots z_{n-2}-z_{i}\cdots z_{n-2}\right) \nonumber\\
&  =-\underset{i<j}{\sum}\frac{k_{i}\cdot k_{j}}{K}\left[  \ln(z_{j}\cdots
z_{n-2})+\ln\left(  1-z_{i}\cdots z_{j-1}\right)  \right]  ,\text{ }%
K=-k_{1}\cdot k_{2},\\
u\left(  x_{i}\right)   &  =\left(  k^{T}\right)  ^{N-2m-q}\underbrace{\left(
k^{L}\right)  ^{2m}}\left(  k^{\prime L}\right)  ^{q}.(k^{\prime L}%
=\frac{\partial k^{L}}{\partial x_{2}}) \label{uuu}%
\end{align}

In Eq.(\ref{uuu}), we have defined%
\begin{equation}
k=\sum_{i\neq2,n}\frac{k_{i}}{x_{i}-x_{2}}=\sum_{i\neq2,n}\frac{k_{i}}%
{z_{i}\cdots z_{n-2}-z_{2}\cdots z_{n-2}}, \label{48}%
\end{equation}
and $k_{\perp}=\left\vert k_{\perp}\right\vert \sum_{i=1}^{r}e^{T_{i}}%
\omega_{i}=\left\vert k_{\perp}\right\vert e^{\hat{T}}$.

The saddle points $\left(  \tilde{z}_{2},\cdots,\tilde{z}_{n-2}\right)  $ are
the solution of\newline%
\begin{equation}
\frac{\partial f}{\partial z_{2}}=0\text{, }\cdots\text{, }\frac{\partial
f}{\partial z_{n-2}}=0. \label{f22}%
\end{equation}
Note that Eq.(\ref{f22}) implies%
\begin{equation}
\tilde{k}^{L}=\frac{\tilde{k}\cdot k_{2}}{M}=\frac{k_{12}}{M}\left.
\frac{\partial f}{\partial x_{2}}\right\vert _{z_{i}=\tilde{z}_{i}}%
=\frac{k_{12}}{M}\left.  \frac{\partial z_{j}}{\partial x_{2}}\frac{\partial
f}{\partial z_{j}}\right\vert _{z_{i}=\tilde{z}_{i}}=0\text{ , }\left\vert
\tilde{k}\right\vert =\left\vert \tilde{k}_{\perp}\right\vert \text{. }
\label{kkk2}%
\end{equation}
We also define%
\begin{equation}
f_{2}\equiv\frac{\partial f}{\partial z_{2}}\text{, }f_{22}\equiv
\frac{\partial^{2}f}{\partial z_{2}^{2}}\text{, }\tilde{f}=f\left(  \tilde
{z}_{2},\cdots,\tilde{z}_{n-2}\right)  \text{, }\tilde{f}_{22}=\left.
\frac{\partial^{2}f}{\partial z_{2}^{2}}\right\vert _{\left(  \tilde{z}%
_{2},\cdots,\tilde{z}_{n-2}\right)  }.
\end{equation}

In view of Eq.(\ref{uuu}) and Eq.(\ref{kkk2}), all up to $(2m)$\textbf{-order
differentiations of }$u$\textbf{ function} in Eq.(\ref{uuu}) at the saddle
point vanish except%
\begin{align}
\left.  \frac{\partial^{2m}u}{\partial z_{2}^{2m}}\right\vert _{\left(
\tilde{z}_{2},\cdots,\tilde{z}_{n-2}\right)  }  &  =\left(  \frac{k_{12}}%
{M}\right)  ^{2m+q}\left(  -\sum_{i\neq2,n}\frac{k_{i}^{T}}{\tilde{x}%
_{i}-\tilde{x}_{2}}\right)  ^{N-2m-2q}\left(  2m\right)  !\left(  \tilde
{f}_{22}\right)  ^{q+2m}\nonumber\\
&  =\left(  \frac{k_{12}}{M}\right)  ^{2m+q}\left(  \tilde{k}^{T}\right)
^{N-2m-2q}\left(  2m\right)  !\left(  \tilde{f}_{22}\right)  ^{q+2m}%
\end{align}

Finally, with the saddle point, we can calculate the $HSSA$ to be%
\begin{align}
\mathcal{T}^{\left(  N,2m,2q\right)  }  &  =\left(  \prod_{i=3}^{n-2}\int%
_{0}^{1}dz_{i}\text{ }z_{i}^{i-2-N}\right)  \int_{0}^{1}dz_{2}\left(
\frac{\partial^{2m}\tilde{u}}{\partial z_{2}^{2m}}\frac{\left(  z_{2}%
-\tilde{z}_{2}\right)  ^{2m}}{\left(  2m\right)  !}\right)  e^{-Kf}\\
&  \simeq\frac{1}{\left(  2m\right)  !}\frac{\partial^{2m}\tilde{u}}{\partial
z_{2}^{2m}}\left(  \prod_{i=3}^{n-2}\tilde{z}_{i}^{i-2-N}\right)  \int_{0}%
^{1}dz_{2}\left(  z_{2}-\tilde{z}_{2}\right)  ^{2m}e^{-Kf\left(  z_{2}\right)
}\\
&  \simeq\frac{1}{\left(  2m\right)  !}\frac{\partial^{2m}\tilde{u}}{\partial
z_{2}^{2m}}\left(  \prod_{i=3}^{n-2}\tilde{z}_{i}^{i-2-N}\right)  \int%
_{0}^{\infty}dz_{2}\left(  z_{2}-\tilde{z}_{2}\right)  ^{2m}e^{-Kf\left(
z_{2}\right)  }\\
&  =\frac{2\sqrt{\pi}}{m!}\left(  \prod_{i=3}^{n-2}\tilde{z}_{i}%
^{i-2-N}\right)  \frac{e^{-K\tilde{f}}}{\left\vert \tilde{k}\right\vert
^{2m+1}}\left.  \frac{\partial^{2m}u}{\partial z_{2}^{2m}}\right\vert
_{z_{i}=\tilde{z}_{i}}\\
&  =2\sqrt{\pi}e^{-K\tilde{f}}\left\vert \tilde{k}\right\vert ^{N-1}\left(
\prod_{i=3}^{n-2}\tilde{z}_{i}^{i-2-N}\right)  \frac{\left(  2m\right)  !}%
{m!}\left(  \frac{-1}{2M}\right)  ^{2m+q}\left(  \frac{2K\tilde{f}_{22}%
}{\left(  \sum_{i\neq2,n}\frac{k_{i}^{T}}{\tilde{x}_{i}-\tilde{x}_{2}}\right)
^{2}}\right)  ^{m+q}%
\end{align}
where $f\left(  z_{2}\right)  =f\left(  z_{2},\tilde{z}_{3},\cdots,\tilde
{z}_{n-2}\right)  $.

The ratios of $n$-point $HSSA$ with $r=1$ is%
\begin{align}
\frac{\mathcal{T}^{\left(  N,m,q\right)  }}{\mathcal{T}^{\left(  N,0,0\right)
}}  &  =\frac{\left(  2m\right)  !}{m!}\left(  \frac{-1}{2M}\right)
^{2m+q}\left(  \frac{2K\tilde{f}_{22}}{\left(  \sum_{i\neq2,n}\frac{k_{i}^{T}%
}{\tilde{x}_{i}-\tilde{x}_{2}}\right)  ^{2}}\right)  ^{m+q}\\
&  =\frac{\left(  2m\right)  !}{m!}\left(  \frac{-1}{2M}\right)  ^{2m+q}%
\end{align}
where the second equality followed from the calculation of decoupling of ZNS.

This suggests the\textbf{ identity}%
\begin{equation}
\frac{2K\tilde{f}_{22}}{\left(  \sum_{i\neq2,n}\frac{k_{i}^{T}}{\tilde{x}%
_{i}-\tilde{x}_{2}}\right)  ^{2}}=1.
\end{equation}
For the case of $n=4$, one can easily solve the saddle point $\tilde{z}%
_{2}=\sec^{2}\frac{\phi}{2}$ to verify the identity.

We have also proved the identity for $n=5$ (complicated to solve saddle
points)%
\begin{equation}
\frac{2K\tilde{f}_{22}}{\left(  \frac{k_{3}^{T}}{\tilde{x}_{3}-\tilde{x}_{2}%
}+\frac{k_{4}^{T}}{1-\tilde{x}_{2}}\right)  ^{2}}=1
\end{equation}
by using \textbf{maple}.

Similar proof can be done for $n=6$%
\begin{equation}
\frac{2K\tilde{f}_{22}}{\left(  \frac{k_{3}^{T}}{\tilde{x}_{3}-\tilde{x}_{2}%
}+\frac{k_{4}^{T}}{\tilde{x}_{4}-\tilde{x}_{2}}+\frac{k_{5}^{T}}{1-\tilde
{x}_{2}}\right)  ^{2}}=1
\end{equation}

\subsection{The $n$-point $HSSA$ with $r=2$}

Now we want to calculate the case of $n$-point $HSSA$ with $r=2$.

We want to calculate $n$-point $HSSA$ with $(n-1)$ tachyons and $1$ high
energy state%
\begin{equation}
\left(  \alpha_{-1}^{T_{1}}\right)  ^{N+p_{1}}\left(  \alpha_{-1}^{T_{2}%
}\right)  ^{p_{2}}\left(  \alpha_{-1}^{L}\right)  ^{2m}\left(  \alpha_{-2}%
^{L}\right)  ^{q}\left\vert 0;k\right\rangle \text{, \ }p_{1}+p_{2}=-2(m+q).
\end{equation}
The ratios of $n$-point $HSSA$ with $r=2$ can be similarly calculated to be%
\begin{align}
\frac{\mathcal{T}^{\left(  p_{1},p_{2},m,q\right)  }}{\mathcal{T}^{\left(
N,0,0,0\right)  }}  &  =\frac{\left(  2m\right)  !}{m!}\left(  \frac{-1}%
{2M}\right)  ^{2m+q}\frac{\left(  2K\tilde{f}_{22}\right)  ^{m+q}}{\left(
\sum_{i\neq2,n}\frac{k_{i}^{T_{1}}}{\tilde{x}_{i}-\tilde{x}_{2}}\right)
^{2m+2q+p_{2}}\left(  \sum_{i\neq2,n}\frac{k_{i}^{T_{2}}}{\tilde{x}_{i}%
-\tilde{x}_{2}}\right)  ^{-p_{2}}}\nonumber\\
&  =\frac{\left(  2m\right)  !}{m!}\left(  \frac{-1}{2M}\right)  ^{2m+q}%
\frac{\left(  \frac{\sum_{i\neq2,n}\frac{k_{i}^{T_{2}}}{\tilde{x}_{i}%
-\tilde{x}_{2}}}{\sum_{i\neq2,n}\frac{k_{i}^{T_{1}}}{\tilde{x}_{i}-\tilde
{x}_{2}}}\right)  ^{p_{2}}}{\left(  \frac{\sum_{i\neq2,n}\frac{k_{i}^{T_{1}}%
}{\tilde{x}_{i}-\tilde{x}_{2}}}{\sqrt{2K\tilde{f}_{22}}}\right)  ^{2m+2q}}
\label{a2}%
\end{align}
On the other hand, the decoupling of ZNS gives%
\begin{equation}
\frac{\mathcal{T}^{\left(  p_{1},p_{2},m,q\right)  }}{\mathcal{T}^{\left(
N,0,0,0\right)  }}=\frac{\left(  2m\right)  !}{m!}\left(  \frac{-1}%
{2M}\right)  ^{2m+q}\omega_{1}^{p_{1}}\omega_{2}^{p_{2}}=\frac{\left(
2m\right)  !}{m!}\left(  \frac{-1}{2M}\right)  ^{2m+q}\frac{(\tan
\theta)^{p_{2}}}{(\cos\theta)^{2m+2q}}. \label{b2}%
\end{equation}
Eq.(\ref{a2}) and Eq.(\ref{b2}) can be \textbf{identified} for any $p_{2}$,
$m$ and $q$ if%
\begin{equation}
\left(  \sum_{i\neq2,n}\frac{k_{i}^{T_{1}}}{\tilde{x}_{i}-\tilde{x}_{2}%
}\right)  =\sqrt{2K\tilde{f}_{22}}\cos\theta\text{, }\left(  \sum_{i\neq
2,n}\frac{k_{i}^{T_{2}}}{\tilde{x}_{i}-\tilde{x}_{2}}\right)  =\sqrt
{2K\tilde{f}_{22}}\sin\theta,
\end{equation}
which implies the identity%
\begin{equation}
\left(  \sum_{i\neq2,n}\frac{k_{i}^{T_{1}}}{\tilde{x}_{i}-\tilde{x}_{2}%
}\right)  ^{2}+\left(  \sum_{i\neq2,n}\frac{k_{i}^{T_{2}}}{\tilde{x}%
_{i}-\tilde{x}_{2}}\right)  ^{2}=2K\tilde{f}_{22}. \label{qq2}%
\end{equation}

\subsection{The $n$-point $HSSA$ with $r\leq24$}

It is now easy to generalize Eq.(\ref{qq2}) to any $r$ (number of $T_{i}$)
with $r\leq24$%
\begin{equation}
\left(  \sum_{i\neq2,n}\frac{k_{i}^{T_{1}}}{\tilde{x}_{i}-\tilde{x}_{2}%
}\right)  ^{2}+\left(  \sum_{i\neq2,n}\frac{k_{i}^{T_{2}}}{\tilde{x}%
_{i}-\tilde{x}_{2}}\right)  ^{2}+\cdots+\left(  \sum_{i\neq2,n}\frac
{k_{i}^{T_{r}}}{\tilde{x}_{i}-\tilde{x}_{2}}\right)  ^{2}=2K\tilde{f}_{22}.
\label{id2}%
\end{equation}
By using Eq.(\ref{48}) and Eq.(\ref{kkk2}), we see that the \textbf{key
identity} Eq.(\ref{id2}) can be written as%
\begin{equation}
\tilde{k}^{2}+2M\tilde{k}^{\prime L}=0. \label{iden2}%
\end{equation}
The ratios in Eq.(\ref{100}) are thus \textbf{proved by the saddle point
method}.

Note that for $n=4$, one can easily solve the saddle point $\tilde{z}_{2}%
=\sec^{2}\frac{\varphi}{2}$ to prove Eq.(\ref{iden2}) analytically. For the
cases of $n=5$ and $6$, we are able to prove the identity numerically. Indeed,
we can see that the identity is a result of the calculation of decoupling of
ZNS. Eq.(\ref{iden2}) is crucial to show the stringy scaling behavior of
$HSSA$.

On the other hand, Eq.(\ref{iden2}) can be used to express $\omega_{i}$ (or
$\theta_{i}$) defined in Eq.(\ref{ww}) in terms of independent kinematics
variables defined in Eq.(\ref{k22}) if one can analytically solve the saddle
point $\tilde{z}_{i}=\left(  \tilde{z}_{2},\cdots,\tilde{z}_{n-2}\right)  $.
Unfortunately, it turns out to be nontrivial except for the case of $n=4$. For
$n=6$ and $r=3$ as an example, the ratios of $6$-point $HSSA$ with $r=3$ can
be calculated to be%
\begin{align}
\frac{\mathcal{T}^{\left(  p_{1},p_{2},p_{3},m,q\right)  }}{\mathcal{T}%
^{\left(  N,0,0,0\right)  }}  &  =\frac{\left(  2m\right)  !}{m!}\left(
\frac{-1}{2M}\right)  ^{2m+q}\frac{\left(  2K\tilde{f}_{22}\right)  ^{m+q}%
}{\left(  \sum_{i\neq2,n=6}\frac{k_{i}^{T_{1}}}{\tilde{x}_{i}-\tilde{x}_{2}%
}\right)  ^{2m+2q+p_{2}+p_{3}}\left(  \sum_{i\neq2,n=6}\frac{k_{i}^{T_{2}}%
}{\tilde{x}_{i}-\tilde{x}_{2}}\right)  ^{-p_{2}}\left(  \sum_{i\neq2,n=6}%
\frac{k_{i}^{T_{3}}}{\tilde{x}_{i}-\tilde{x}_{2}}\right)  ^{-p_{3}}%
}\nonumber\\
&  =\frac{\left(  2m\right)  !}{m!}\left(  \frac{-1}{2M}\right)  ^{2m+q}%
\frac{\left(  \frac{\sum_{i\neq2,n=6}\frac{k_{i}^{T_{2}}}{\tilde{x}_{i}%
-\tilde{x}_{2}}}{\sum_{i\neq2,n=6}\frac{k_{i}^{T_{1}}}{\tilde{x}_{i}-\tilde
{x}_{2}}}\right)  ^{p_{2}}\left(  \frac{\sum_{i\neq2,n=6}\frac{k_{i}^{T_{3}}%
}{\tilde{x}_{i}-\tilde{x}_{2}}}{\sum_{i\neq2,n=6}\frac{k_{i}^{T_{1}}}%
{\tilde{x}_{i}-\tilde{x}_{2}}}\right)  ^{p_{3}}}{\left(  \frac{\sum_{i\neq
2,n}\frac{k_{i}^{T_{1}}}{\tilde{x}_{i}-\tilde{x}_{2}}}{\sqrt{2K\tilde{f}_{22}%
}}\right)  ^{2m+2q}}. \label{a22}%
\end{align}
On the other hand, the decoupling of ZNS gives%
\begin{equation}
\frac{\mathcal{T}^{\left(  p_{1},p_{2},m,q\right)  }}{\mathcal{T}^{\left(
N,0,0,0\right)  }}=\frac{\left(  2m\right)  !}{m!}\left(  \frac{-1}%
{2M}\right)  ^{2m+q}\omega_{1}^{p_{1}}\omega_{2}^{p_{2}}\omega_{3}^{p3}%
=\frac{\left(  2m\right)  !}{m!}\left(  \frac{-1}{2M}\right)  ^{2m+q}%
\frac{(\tan\theta_{1}\cos\theta_{2})^{p_{2}}(\tan\theta_{1}\sin\theta
_{2})^{p_{3}}}{(\cos\theta_{1})^{2m+2q}}. \label{b22}%
\end{equation}
Eq.(\ref{a22}) and Eq.(\ref{b22}) can be \textbf{identified} for any $p_{2}$,
$p_{3}$, $m$ and $q$ if%
\begin{align}
\sum_{i\neq2,n=6}\frac{k_{i}^{T_{1}}}{\tilde{x}_{i}-\tilde{x}_{2}}  &
=\sqrt{2K\tilde{f}_{22}}\cos\theta_{1}\text{,}\label{1a}\\
\text{ }\frac{\text{ }\sum_{i\neq2,n=6}\frac{k_{i}^{T_{2}}}{\tilde{x}%
_{i}-\tilde{x}_{2}}}{\text{ }\sum_{i\neq2,n=6}\frac{k_{i}^{T_{1}}}{\tilde
{x}_{i}-\tilde{x}_{2}}}  &  =\sqrt{2K\tilde{f}_{22}}\tan\theta_{1}\cos
\theta_{2},\label{2b}\\
\frac{\text{ }\sum_{i\neq2,n=6}\frac{k_{i}^{T_{3}}}{\tilde{x}_{i}-\tilde
{x}_{2}}}{\text{ }\sum_{i\neq2,n=6}\frac{k_{i}^{T_{1}}}{\tilde{x}_{i}%
-\tilde{x}_{2}}}  &  =\sqrt{2K\tilde{f}_{22}}\tan\theta_{1}\sin\theta_{2},
\label{3c}%
\end{align}
which allow us to formally express $\theta_{1}$ and $\theta_{2}$ in terms of
kinematics variables defined in Eq.(\ref{k22}) as%
\begin{equation}
\theta_{1}=\arctan\frac{\sqrt{\left(  \tilde{k}_{\bot}^{T_{2}}\right)
^{2}+\left(  \tilde{k}_{\bot}^{T_{3}}\right)  ^{2}}}{\tilde{k}_{\perp}^{T_{1}%
}}\text{, }\theta_{2}=\arctan\frac{\tilde{k}_{\perp}^{T_{3}}}{\tilde{k}%
_{\perp}^{T_{2}}}. \label{RG}%
\end{equation}

It is important to note from Eq.(\ref{16}) that the ratios of $6$-point $HSSA$
with $r=3$ depends only on $2$ kinematics variables $\theta_{1}$ and
$\theta_{2}$ instead of $8$. We see that Eq.(\ref{1a}), Eq.(\ref{2b}) and
Eq.(\ref{3c}) gives the key identity in Eq.(\ref{id2}) for the case of $n=6$
and $r=3$.

\section{Conclusion}

\subsection{Bjorken Scaling}

These stringy scaling behaviors are reminiscent of deep \textbf{inelastic
scattering of electron and proton} where the \textbf{two structure functions}
$W_{1}(Q^{2},\nu)$ and $W_{2}(Q^{2},\nu)$ scale, and become not functions of
$2$ kinematics variables $Q^{2}$ and $\nu$ independently but only of their
ratio $Q^{2}/\nu$. The number of independent kinematics variables
\textbf{reduces} from $2$ to $1$.

That is, the structure functions scale as%
\begin{equation}
MW_{1}(Q^{2},\nu)\rightarrow F_{1}(x),\text{ \ \ }\nu W_{2}(Q^{2}%
,\nu)\rightarrow F_{2}(x)
\end{equation}
where $x$ is the \textbf{Bjorken variable} and $M$ is the proton mass.

Moreover, due to the spin-$\frac{1}{2}$ assumption of quark, \textbf{Callan
and Gross} derived the relation%
\begin{equation}
2xF_{1}(x)=F_{2}(x).
\end{equation}

\subsection{\textit{Scaling relations }among critical exponents}

These stringy scaling behaviors also remind us of the \textbf{scaling
relations}\textit{ }among critical exponents through \textbf{Wilson's RG}
analysis of \textbf{Widom} hypothesis in \textbf{statistical mechanics.}

There a set of $6$ exponents $\alpha$, $\beta$, $\gamma$, $\delta$, $\nu$ and
$\eta$ are related by $4$ scaling relations, and the number of independent
exponents \textbf{reduces} from $6$ to $2$.

\section{Stringy scaling of Regge string scattering amplitudes}

A class of $n$-point \textbf{Regge} string scattering amplitudes shows similar
scaling behaviors. All results have been published in \cite{hard,Regge}.

\end{document}